\documentclass[sigconf]{acmart}

\usepackage{booktabs}
\usepackage{cleveref}
\usepackage{dirtytalk}
\usepackage{natbib}
\usepackage{subfig}
\usepackage{enumitem}
\usepackage{url}
\usepackage{bbding} 
\usepackage{framed}

\begin{document}

\title{The Importance of Conflict Resolution Techniques in Autonomous Agile Teams}

\author{Lucas Gren and Per Lenberg}
\orcid{1234-5678-9012}
\affiliation{%
  \institution{Chalmers University of Technology and The University of Gothenburg\\}
  \streetaddress{The Department of Computer Science and Engineering}
  \city{Gothenburg} 
  \state{Sweden} 
  \postcode{412--92}
}
\email{lucas.gren@cse.gu.se | perle@chalmers.se}

\begin{abstract}
Today, software companies usually organize their work in teams. Social science research on team development has shown that for a team to reach a productive and autonomous stage, it has to be able to manage internal conflicts and disagreements efficiently. To better facilitate the team development process, we argue that software engineers needs additional training in negotiation skills and conflict resolution. In this position paper, we outline ideas for what aspects to consider in such training. As an example, we argue that a majority of the conflicts originate from team-level factors and that they, therefore, should be managed on the team-level instead of in relation to dyads.

\end{abstract}


\copyrightyear{2018} 
\acmYear{2018} 
\setcopyright{acmlicensed}
\acmConference[XP '18 Companion]{XP '18 Companion}{May 21--25, 2018}{Porto, Portugal}
\acmPrice{15.00}
\acmDOI{10.1145/3234152.3234185}
\acmISBN{978-1-4503-6422-5/18/05}


\maketitle

\section{Introduction}\label{sec:introduction}
The introduction of the agile methods has shifted the focus from the individual software developer and instead highlighted team, collaboration, and communication \citep{cockburn2001agilepeoplefactor}. In software engineering organizations today, well-functioning teams are considered a critical success factor \citep{grenjss2}. A natural consequence, or a byproduct, of increased collaboration is interpersonal conflict \citep{gren2017links}.

To obtain well-functioning and autonomous teams, a set of group psychological factors has to be in place. Self-organization of teams has been shown to surface naturally only in the more mature stages of group development, which also implies that the leadership gets more and more shared over time, and many groups do not reach the more mature stages but get instead stuck for a variety of reasons \citep{wheelan2003}. The group developmental theories state that, when humans organize in small groups to achieve a set of common goals, we go through a specific set of stages and the group members behave differently across these stages \citep{kozlowski2006enhancing}. 

Research on development of small groups agrees on that a period of disagreement and conflict is necessary to reach the better functioning mature stages \citep{kozlowski2006enhancing}. People in groups need to challenge one another to figure out the group members' real competences and, also, set the group norms, i.e., the rules of the game \citep{wheelandev}. This implies that some conflict stage is needed for most teams in order to later be more effective, and teams need to create a practical conflict management approach specific for every single constellation of people. Having efficient conflict resolution techniques in agile teams are thus a prerequisite for building a well functioning autonomous team. Therefore, conflict resolution needs to be conducted on team level, which has also been shown in the software engineering context in a study by \citet{ocker2001relationship}. They showed that the group development maturity was positively connected to the quality of work output, and the degree of satisfaction.

In this short paper, we first outline research on work-related conflicts from the information systems and software engineering domain. We then present guidelines taken from conflict resolution research and, finally, we discuss potential gains in the software engineering autonomous teams' context and suggest future work. 

\section{Interpersonal Conflict and Software Engineering Research}
Traditionally, psychology researchers divide conflicts into the three types (relation, process, and task) based on their content. Still, these types are not well-defined and their link to performance not fully understood \citep{behfar2008critical}. As an example, relationship conflicts have been shown to affect both task-based and social aspects of team performance negatively \citep{manata2016exploring}. Therefore, there seem to be indications of more complex relationships between conflict types than presented by, for example, \citet{domino2003conflict} within the software development domain.

A conflict can, in its broader sense, be defined as ``the process which begins when one party perceives that another has frustrated, or is about to frustrate, some concern of hers or his'' \citep{thomas1992conflict}. Therefore, a conflict has nothing to do with raising one's voice of fighting, even if that is the practical interpretation of the word in some languages, like Swedish.

Information system (IS) researchers have also conducted studies related to conflict. In a study from 2001, \citet{barki2001interpersonal} showed that interpersonal conflict consisting of disagreement, interference, and negative emotion had less of an impact on the project outcomes when the teams had well-functioning conflict management \citep{barki2001interpersonal}. Similar results were obtained in that same year by \citet{sawyer2001effects}.

The research on conflict in software engineering is scarce, which might indicate the difficulty of such inquiries. Among the older studies is the work by \citet{gobeli1998managing} where they show that dysfunctional conflict management approaches have adverse effects on results. In a study on requirements specification, interpersonal conflicts were shown to link directly to requirements diversity, which was negatively associated to project performance \citep{liu2011relationships}. Furthermore, a study by \citet{gren2017links} showed that interpersonal conflict was adversely connected to the agile team practices Iterative Development and Customer Access.

Together, these mentioned studies further motivate the need for proper conflict resolution in agile teams. Therefore, in the following sections, we will present techniques for how software organizations can raise the knowledge of having to manage interpersonal conflict efficiently.

\section{Intra- and inter-group conflict}
Interpersonal conflict manifests itself often i dyadic relations. A work- or relationship-related conflict needs to be verbally expressed by one person at the time and most often directed to another individual. However, this does not mean that the conflict is isolated to the individuals expressing it \citep{wall1995conflict}. In fact conflicts are seen to be between two parties, be it in individuals, groups or nations \citep{pruitt1969stability}.

Intra-team conflicts, we argue, need a structure to be managed at an early point in time, since conflicts are known to escalate, and sometimes quite severely over time \citep{pruitt1969stability}. Therefore, teams are helped by discussing early conflicts continuously before they become infected and personal. However, if a conflict has escalated, there are expensive knowledge on how to behave in order to solve conflicts fast depending on personal stake, rhetoric, etc. Even is the section below focuses on individuals they techniques can be extended to any two parties \citep{pruitt1969stability}.

\section{Escalated interpersonal conflict}
In this section, we summarize the content from a number of practical handbooks on conflict management, since we want to provide hands-on tips of how to reason around conflict. It is intended as an introduction to conflict management in practice. For an extensive review of conflict resolution research, we recommend \citet{coleman2014tho} that includes almost a thousand pages and hundreds of references to academic papers. We would, again, like to highlight that the conflict resolution needs to be on team-level since they are a prerequisite for getting a team to mature over time.

There is a diversity of situations that potentially can lead to interpersonal team conflict. For example competing needs, fighting about scarce resources, misunderstandings, unclear situations, different views on roles or divisions, different values, norms or understandings, communication problems, competition\slash rivalry, organizational change, and stress \citep{coleman2014tho,wall1995conflict}. 

Having high emotional intelligence is a very useful for successful conflict management. Below is a list of common mistakes that are known to trigger aggressive or unwilling responses in conflict situations \citep{goleman1998wwe,wall1995conflict}:

\begin{itemize}
\item One perspective --- To see the problem only from your perspective.
\item Poor communication --- To stop listening\slash understanding.
\item Only binary options --- Think ``right or wrong;'' there's only one way, and that's my way.
\item Correspondence bias --- It's not just the concrete issue that is the problem, it's the person.
\item Add new information --- Bringing up new information not know to the other party. 
\item Manipulation --- Withholding information, talk behind people's backs.
\item Hurting purposefully --- Finding personal weak spots and attacking.
\item Ignoring social rules --- Stop saying hello, ignore, and exclude from mailing lists.
\end{itemize}
If successfully avoiding the above mentioned mistakes, and instead recognizing other people's perspectives and referring to one's own role in the conflict, trigger much more willingness to find agreeable solutions:

\begin{itemize}
\item I-message (not iMessage) --- Meaning that arguments are more effective if they refer to the person talking instead of the person referring to a set of people or groups not present. Conflict should also be resolved, as a first step, in private using face-to-face communication.
\item Speak about what you want yourself, not what the other one ``should'' want. Describe your problem with the other person's action\slash behavior and not personality. Listen to the other person and show that you understand the content of what the person is saying. One way of easily achieving this is to verbally interpret what the other person just said, e.g.\ ``if I understand you correctly you mean that...''
\item Define the problem as a mutual, narrow and specific problem. 
\item Describe your feelings connected to the problem (sad, angry, frustrated, disrespected etc.)
\item Exchange motives to your positions, what's behind your different views? What needs to be fulfilled? Listen to each others' perspectives.
\item Identify possibilities for mutual benefit by providing many possible solutions, and chose one wisely \citep{wall1995conflict}.
\end{itemize}

A clearer step-by-step protocol might be the following: 

\begin{itemize}
\item A: Now (What's the present situation? This is what I\slash we\slash they do now)
\item B: Desired end result (This is how I want it. I\slash we\slash they should do like this).
\item C: Obstacles (Why A instead of B?).
\item C1: Do we know about the obstacles?
\item C2: Are the obstacles possible to remove?
\item C3: Can we remove the obstacles?
\item C4: Do we want to remove the obstacles?
\item D: Actions (Suggestions\slash changes) \citep{wall1995conflict}.
\end{itemize}

It is important to recognize that different approaches are needed depending on how infected the conflicts are. One significant intervention when conflicts are more complicated is to use a mediator \citep{moore2003tmp}. In the agile software development context, the process facilitator (i.e., the Scrum Master in Scrum) would be ideal to take on such a role when needed. \citet{grenjss2} also showed that Scrum Masters often do manage teams in such a way in practice. We recognize that such behavior is not considered to be ``pure Scrum,'' but argue for the usefulness of having a formal protocol for how agile teams should manage conflict step-by-step in software development organizations.

\begin{figure*}
\centerline{\includegraphics[scale=0.8]{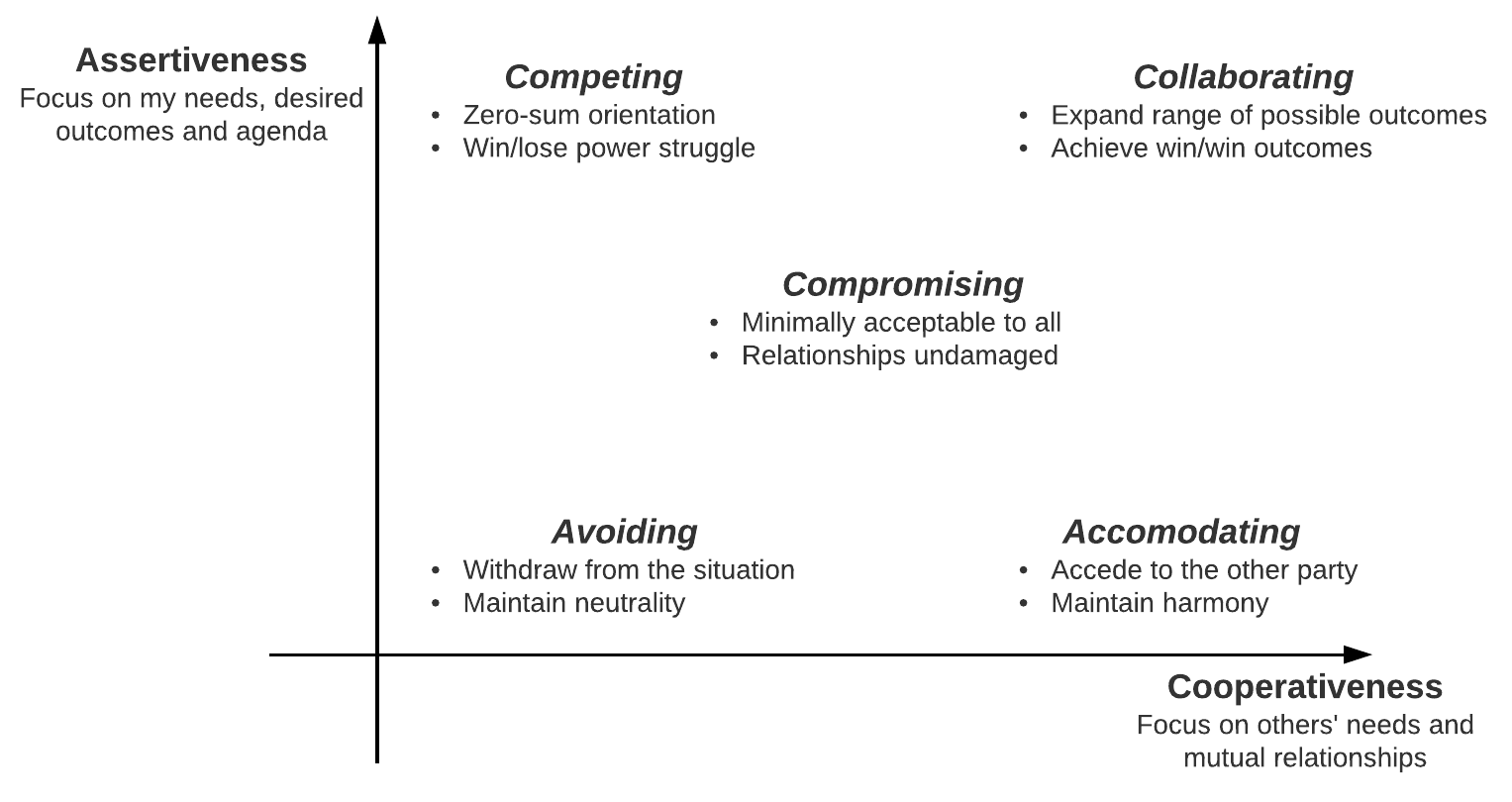}}
\caption{The Thomas-Kilmann Conflict Modes (adopted from \cite{thomas1992conflict}).}
\label{kilmann}
\end{figure*}

It is also important to acknowledge that employees have disparate interests in different conflicts. A well-used model of such stance in conflicts was suggested by \citet{thomas1992conflict}, and is shown in Figure~\ref{kilmann}. Depending on the assertiveness and cooperativeness in each conflict a person will approach the conflict mainly in five different ways (although people tend to resort to some of them more than others). With low assertiveness, i.e., focus on own needs, and low cooperativeness the person will avoid the conflict and maintain their neutrality in relation to the conflict. With high assertiveness and low cooperativeness the person participated by having a zero-sum orientation and assumes that one has to win and the other has to lose. With high cooperativeness but low assertiveness, the person maintain harmony and accede to the other party. With an intermediate level on both assertiveness and cooperativeness, the person will compromise and try to find solutions acceptable to all parties, which also maintains the relationship undamaged. With high levels of both assertiveness and cooperativeness, the person will collaborate, meaning that the person will try to expand the range of possible outcomes and achieve win\slash win outcomes, which also challenges the relationship more.

There is also a range of cognitive biases that might create conflict that could be avoided (for more examples of such cognitive biases see, e.g., \citet{evans1989bih}). The literature on cognitive and biases is vast, and we will only mention one of them in this paper. The one we have chosen that we believe have a significant impact on conflict resolution is the \emph{correspondence bias} already mentioned in the list above about common mistakes in conflict situations. This bias is known by many names and was first called the \emph{fundamental attribution error}. This error is about people's tendency to place an undue emphasis on internal characteristics to explain the behavior of someone else in a given situation, rather than considering external factors, i.e., the tendency to believe that people’s actions reflect who they are \citep{gawronski2004theory}. Therefore, when observing an inappropriate behavior, it is critical to take into account and recognize the situational factors, i.e., not only resort to individual factors, such as personality, to explain the behavior (see \citet{coleman2014tho} pp.\ 502). This further motivates avoiding to turn team-level problems into personal ones.

\section{Discussion}\label{sec:discussion}
The human factors are increasingly being recognized by software engineering researchers and practitioners alike \citep{lenberg2015behavioral}. The psychological and sociological aspects have been presented as the missing piece in software engineering education \citep{seovercoming}. Such approaches to training already exist in other fields and can directly be applied to the software engineering context (see e.g., \citet{shell2001teaching}).

In the agile manifesto \citep{fowler2001}, the value ``Individuals and interactions over processes and tools'' emphasizes the importance of making human interaction efficient. A core aspect of such interactions is the ability to manage conflict well. In order to build efficient and autonomous teams in software organizations, having a formal structure for conflict resolution would undoubtedly be helpful. Research conducted in the information system (IS) domain has shown that making employees aware of how conflicts work has positive effects \citep{barki2001interpersonal}.

To increase awareness and to raise software organizations' general understanding of interpersonal conflicts, we suggest that software engineering education should include negotiation and conflict resolution training \citep{shell2001teaching}. Since software engineers tend to conduct their work in small groups, we suggest that such training should emphasize the group aspects, i.e., interpersonal conflict in autonomous agile teams should be seen a group-level problem and not as a dyadic problem. We believe a majority of conflicts are not due to individual factors but instead team-related contextual factors such as poor communication, unclear role, or undefined goals. These dissimilarities can, therefore, not be resolved through addressing the individuals involved only, but should instead be managed on a team-level.

As mentioned in the previous section, organizations need to provide agile teams with a well-defined process of how to manage team conflict in the organization, and the Scrum Master role might be appropriate for facilitating this process. If such guidelines are not in place, it will be cumbersome to trust team with the authority they need to set directions for new products.

\section{Conclusions and future work}\label{sec:future}
In this position paper, we emphasize the importance of considering conflicts in software organizations. Social science research on group dynamics and team development have repeatedly shown that for a team to reach a productive stage it has to, in an efficient way, be able to manage internal conflicts and disagreements. To increase the software engineering general knowledge on how to handle disagreements within a team, we also suggest that software engineering education should include negotiation and conflict resolution training. In this papers, we have provided initial ideas for what aspects to consider in such training. As an example, we argue that a majority of the conflicts originate from group-related factors and that they, therefore, should be managed using a team-level approach.

\bibliographystyle{ACM-Reference-Format}
\bibliography{references}  


\begin{thebibliography}{00}


\ifx \showCODEN    \undefined \def \showCODEN     #1{\unskip}     \fi
\ifx \showDOI      \undefined \def \showDOI       #1{{\tt DOI:}\penalty0{#1}\ }
  \fi
\ifx \showISBNx    \undefined \def \showISBNx     #1{\unskip}     \fi
\ifx \showISBNxiii \undefined \def \showISBNxiii  #1{\unskip}     \fi
\ifx \showISSN     \undefined \def \showISSN      #1{\unskip}     \fi
\ifx \showLCCN     \undefined \def \showLCCN      #1{\unskip}     \fi
\ifx \shownote     \undefined \def \shownote      #1{#1}          \fi
\ifx \showarticletitle \undefined \def \showarticletitle #1{#1}   \fi
\ifx \showURL      \undefined \def \showURL       #1{#1}          \fi
\providecommand\bibfield[2]{#2}
\providecommand\bibinfo[2]{#2}
\providecommand\natexlab[1]{#1}
\providecommand\showeprint[2][]{arXiv:#2}

\bibitem[\protect\citeauthoryear{Barki and Hartwick}{Barki and
  Hartwick}{2001}]%
        {barki2001interpersonal}
\bibfield{author}{\bibinfo{person}{Henri Barki} {and} \bibinfo{person}{Jon
  Hartwick}.} \bibinfo{year}{2001}\natexlab{}.
\newblock \showarticletitle{Interpersonal conflict and its management in
  information system development}.
\newblock \bibinfo{journal}{{\em MIS Quarterly\/}} (\bibinfo{year}{2001}),
  \bibinfo{pages}{195--228}.
\newblock


\bibitem[\protect\citeauthoryear{Behfar, Peterson, Mannix, and Trochim}{Behfar
  et~al\mbox{.}}{2008}]%
        {behfar2008critical}
\bibfield{author}{\bibinfo{person}{Kristin~J Behfar},
  \bibinfo{person}{Randall~S Peterson}, \bibinfo{person}{Elizabeth~A Mannix},
  {and} \bibinfo{person}{William~MK Trochim}.} \bibinfo{year}{2008}\natexlab{}.
\newblock \showarticletitle{The critical role of conflict resolution in teams:
  A close look at the links between conflict type, conflict management
  strategies, and team outcomes.}
\newblock \bibinfo{journal}{{\em Journal of applied psychology\/}}
  \bibinfo{volume}{93}, \bibinfo{number}{1} (\bibinfo{year}{2008}),
  \bibinfo{pages}{170}.
\newblock


\bibitem[\protect\citeauthoryear{Cockburn and Highsmith}{Cockburn and
  Highsmith}{2001}]%
        {cockburn2001agilepeoplefactor}
\bibfield{author}{\bibinfo{person}{Alistair Cockburn} {and}
  \bibinfo{person}{Jim Highsmith}.} \bibinfo{year}{2001}\natexlab{}.
\newblock \showarticletitle{Agile software development: {T}he people factor}.
\newblock \bibinfo{journal}{{\em IEEE Computer\/}} \bibinfo{number}{11}
  (\bibinfo{year}{2001}), \bibinfo{pages}{131--133}.
\newblock


\bibitem[\protect\citeauthoryear{Coleman, Deutsch, and Marcus}{Coleman
  et~al\mbox{.}}{2014}]%
        {coleman2014tho}
\bibfield{author}{\bibinfo{person}{Peter~T. Coleman}, \bibinfo{person}{Morton
  Deutsch}, {and} \bibinfo{person}{Eric~C. Marcus}.}
  \bibinfo{year}{2014}\natexlab{}.
\newblock \bibinfo{booktitle}{{\em The Handbook of Conflict Resolution: Theory
  and Practice\/} (\bibinfo{edition}{3rd} ed.)}.
\newblock \bibinfo{publisher}{John Wiley {\&} Sons}, \bibinfo{address}{San
  Francisco, California}.
\newblock


\bibitem[\protect\citeauthoryear{Domino, Collins, Hevner, and Cohen}{Domino
  et~al\mbox{.}}{2003}]%
        {domino2003conflict}
\bibfield{author}{\bibinfo{person}{Madeline~Ann Domino},
  \bibinfo{person}{Rosann~Webb Collins}, \bibinfo{person}{Alan~R Hevner}, {and}
  \bibinfo{person}{Cynthia~F Cohen}.} \bibinfo{year}{2003}\natexlab{}.
\newblock \showarticletitle{Conflict in collaborative software development}. In
  \bibinfo{booktitle}{{\em Proceedings of the 2003 SIGMIS conference on
  Computer personnel research}}. ACM, \bibinfo{pages}{44--51}.
\newblock


\bibitem[\protect\citeauthoryear{Evans}{Evans}{1989}]%
        {evans1989bih}
\bibfield{author}{\bibinfo{person}{Jonathan St. B.~T. Evans}.}
  \bibinfo{year}{1989}\natexlab{}.
\newblock \bibinfo{booktitle}{{\em Bias in human reasoning: Causes and
  consequences}}.
\newblock \bibinfo{publisher}{Erlbaum}, \bibinfo{address}{London}.
\newblock


\bibitem[\protect\citeauthoryear{Fowler and Highsmith}{Fowler and
  Highsmith}{2001}]%
        {fowler2001}
\bibfield{author}{\bibinfo{person}{M. Fowler} {and} \bibinfo{person}{J.
  Highsmith}.} \bibinfo{year}{2001}\natexlab{}.
\newblock \bibinfo{title}{{The Agile Manifesto}}.
\newblock \bibinfo{howpublished}{In Software Development, Issue on Agile
  Methodologies, last accessed on December 29th, 2006}.   (\bibinfo{date}{Aug.}
  \bibinfo{year}{2001}).
\newblock


\bibitem[\protect\citeauthoryear{Gawronski}{Gawronski}{2004}]%
        {gawronski2004theory}
\bibfield{author}{\bibinfo{person}{Bertram Gawronski}.}
  \bibinfo{year}{2004}\natexlab{}.
\newblock \showarticletitle{Theory-based bias correction in dispositional
  inference: The fundamental attribution error is dead, long live the
  correspondence bias}.
\newblock \bibinfo{journal}{{\em European review of social psychology\/}}
  \bibinfo{volume}{15}, \bibinfo{number}{1} (\bibinfo{year}{2004}),
  \bibinfo{pages}{183--217}.
\newblock


\bibitem[\protect\citeauthoryear{Gobeli, Koenig, and Bechinger}{Gobeli
  et~al\mbox{.}}{1998}]%
        {gobeli1998managing}
\bibfield{author}{\bibinfo{person}{David~H Gobeli}, \bibinfo{person}{Harold~F
  Koenig}, {and} \bibinfo{person}{Iris Bechinger}.}
  \bibinfo{year}{1998}\natexlab{}.
\newblock \showarticletitle{Managing conflict in software development teams: A
  multilevel analysis}.
\newblock \bibinfo{journal}{{\em Journal of Product Innovation Management\/}}
  \bibinfo{volume}{15}, \bibinfo{number}{5} (\bibinfo{year}{1998}),
  \bibinfo{pages}{423--435}.
\newblock


\bibitem[\protect\citeauthoryear{Goleman}{Goleman}{1998}]%
        {goleman1998wwe}
\bibfield{author}{\bibinfo{person}{Daniel Goleman}.}
  \bibinfo{year}{1998}\natexlab{}.
\newblock \bibinfo{booktitle}{{\em Working with emotional intelligence}}.
\newblock \bibinfo{publisher}{Bantam Books}, \bibinfo{address}{New York}.
\newblock


\bibitem[\protect\citeauthoryear{Gren}{Gren}{2017}]%
        {gren2017links}
\bibfield{author}{\bibinfo{person}{Lucas Gren}.}
  \bibinfo{year}{2017}\natexlab{}.
\newblock \showarticletitle{The Links Between Agile Practices, Interpersonal
  Conflict, and Perceived Productivity}. In \bibinfo{booktitle}{{\em
  Proceedings of the 21st International Conference on Evaluation and Assessment
  in Software Engineering}}. ACM, \bibinfo{pages}{292--297}.
\newblock


\bibitem[\protect\citeauthoryear{Gren, Torkar, and Feldt}{Gren
  et~al\mbox{.}}{2017}]%
        {grenjss2}
\bibfield{author}{\bibinfo{person}{L Gren}, \bibinfo{person}{R Torkar}, {and}
  \bibinfo{person}{R Feldt}.} \bibinfo{year}{2017}\natexlab{}.
\newblock \showarticletitle{Group development and group maturity when building
  agile teams: {A} qualitative and quantitative investigation at eight large
  companies}.
\newblock \bibinfo{journal}{{\em The Journal of Systems and Software\/}}
  \bibinfo{volume}{124} (\bibinfo{year}{2017}), \bibinfo{pages}{104—--119}.
\newblock
\showDOI{%
\url{http://dx.doi.org/10.1016/j.jss.2016.11.024}}


\bibitem[\protect\citeauthoryear{Kozlowski and Ilgen}{Kozlowski and
  Ilgen}{2006}]%
        {kozlowski2006enhancing}
\bibfield{author}{\bibinfo{person}{Steve~WJ Kozlowski} {and}
  \bibinfo{person}{Daniel~R Ilgen}.} \bibinfo{year}{2006}\natexlab{}.
\newblock \showarticletitle{Enhancing the effectiveness of work groups and
  teams}.
\newblock \bibinfo{journal}{{\em Psychological science in the public
  interest\/}} \bibinfo{volume}{7}, \bibinfo{number}{3} (\bibinfo{year}{2006}),
  \bibinfo{pages}{77--124}.
\newblock


\bibitem[\protect\citeauthoryear{Lenberg, Feldt, and Wallgren}{Lenberg
  et~al\mbox{.}}{2015}]%
        {lenberg2015behavioral}
\bibfield{author}{\bibinfo{person}{Per Lenberg}, \bibinfo{person}{Robert
  Feldt}, {and} \bibinfo{person}{Lars~G{\"o}ran Wallgren}.}
  \bibinfo{year}{2015}\natexlab{}.
\newblock \showarticletitle{Behavioral software engineering: A definition and
  systematic literature review}.
\newblock \bibinfo{journal}{{\em Journal of Systems and software\/}}
  \bibinfo{volume}{107} (\bibinfo{year}{2015}), \bibinfo{pages}{15--37}.
\newblock


\bibitem[\protect\citeauthoryear{Liu, Chen, Chen, and Sheu}{Liu
  et~al\mbox{.}}{2011}]%
        {liu2011relationships}
\bibfield{author}{\bibinfo{person}{Julie Yu-Chih Liu}, \bibinfo{person}{Hun-Gee
  Chen}, \bibinfo{person}{Charlie~C Chen}, {and} \bibinfo{person}{Tsong~Shin
  Sheu}.} \bibinfo{year}{2011}\natexlab{}.
\newblock \showarticletitle{Relationships among interpersonal conflict,
  requirements uncertainty, and software project performance}.
\newblock \bibinfo{journal}{{\em International Journal of Project
  Management\/}} \bibinfo{volume}{29}, \bibinfo{number}{5}
  (\bibinfo{year}{2011}), \bibinfo{pages}{547--556}.
\newblock


\bibitem[\protect\citeauthoryear{Manata}{Manata}{2016}]%
        {manata2016exploring}
\bibfield{author}{\bibinfo{person}{Brian Manata}.}
  \bibinfo{year}{2016}\natexlab{}.
\newblock \showarticletitle{Exploring the association between relationship
  conflict and group performance.}
\newblock \bibinfo{journal}{{\em Group Dynamics: Theory, Research, and
  Practice\/}} \bibinfo{volume}{20}, \bibinfo{number}{2}
  (\bibinfo{year}{2016}), \bibinfo{pages}{93--104}.
\newblock


\bibitem[\protect\citeauthoryear{Moore}{Moore}{2003}]%
        {moore2003tmp}
\bibfield{author}{\bibinfo{person}{Christopher~W. Moore}.}
  \bibinfo{year}{2003}\natexlab{}.
\newblock \bibinfo{booktitle}{{\em The mediation process: Practical strategies
  for resolving conflict\/} (\bibinfo{edition}{3rd} ed.)}.
\newblock \bibinfo{publisher}{Jossey-Bass}, \bibinfo{address}{San Francisco,
  California}.
\newblock


\bibitem[\protect\citeauthoryear{Ocker}{Ocker}{2001}]%
        {ocker2001relationship}
\bibfield{author}{\bibinfo{person}{Rosalie~J Ocker}.}
  \bibinfo{year}{2001}\natexlab{}.
\newblock \showarticletitle{The relationship between interaction, group
  development, and outcome: {A} study of virtual communication}. In
  \bibinfo{booktitle}{{\em Proceedings of the 34th Annual Hawaii International
  Conference on System Sciences}}. IEEE, \bibinfo{pages}{1--10}.
\newblock


\bibitem[\protect\citeauthoryear{Pruitt}{Pruitt}{1969}]%
        {pruitt1969stability}
\bibfield{author}{\bibinfo{person}{Dean~G Pruitt}.}
  \bibinfo{year}{1969}\natexlab{}.
\newblock \showarticletitle{Stability and sudden change in interpersonal and
  international affairs}.
\newblock \bibinfo{journal}{{\em Journal of Conflict Resolution\/}}
  \bibinfo{volume}{13}, \bibinfo{number}{1} (\bibinfo{year}{1969}),
  \bibinfo{pages}{18--38}.
\newblock


\bibitem[\protect\citeauthoryear{Sawyer}{Sawyer}{2001}]%
        {sawyer2001effects}
\bibfield{author}{\bibinfo{person}{Steve Sawyer}.}
  \bibinfo{year}{2001}\natexlab{}.
\newblock \showarticletitle{Effects of intra-group conflict on packaged
  software development team performance}.
\newblock \bibinfo{journal}{{\em Information Systems Journal\/}}
  \bibinfo{volume}{11}, \bibinfo{number}{2} (\bibinfo{year}{2001}),
  \bibinfo{pages}{155--178}.
\newblock


\bibitem[\protect\citeauthoryear{Shell}{Shell}{2001}]%
        {shell2001teaching}
\bibfield{author}{\bibinfo{person}{G~Richard Shell}.}
  \bibinfo{year}{2001}\natexlab{}.
\newblock \showarticletitle{Teaching Ideas: Bargaining Styles and Negotiation:
  The {T}homas-{K}ilmann Conflict Mode Instrument in Negotiation Training}.
\newblock \bibinfo{journal}{{\em Negotiation Journal\/}} \bibinfo{volume}{17},
  \bibinfo{number}{2} (\bibinfo{year}{2001}), \bibinfo{pages}{155--174}.
\newblock


\bibitem[\protect\citeauthoryear{Thomas}{Thomas}{1992}]%
        {thomas1992conflict}
\bibfield{author}{\bibinfo{person}{Kenneth~W Thomas}.}
  \bibinfo{year}{1992}\natexlab{}.
\newblock \showarticletitle{Conflict and conflict management: Reflections and
  update}.
\newblock \bibinfo{journal}{{\em Journal of organizational behavior\/}}
  \bibinfo{volume}{13}, \bibinfo{number}{3} (\bibinfo{year}{1992}),
  \bibinfo{pages}{265--274}.
\newblock


\bibitem[\protect\citeauthoryear{Wall~Jr and Callister}{Wall~Jr and
  Callister}{1995}]%
        {wall1995conflict}
\bibfield{author}{\bibinfo{person}{James~A Wall~Jr} {and}
  \bibinfo{person}{Ronda~Roberts Callister}.} \bibinfo{year}{1995}\natexlab{}.
\newblock \showarticletitle{Conflict and its management}.
\newblock \bibinfo{journal}{{\em Journal of management\/}}
  \bibinfo{volume}{21}, \bibinfo{number}{3} (\bibinfo{year}{1995}),
  \bibinfo{pages}{515--558}.
\newblock


\bibitem[\protect\citeauthoryear{Wheelan}{Wheelan}{2005}]%
        {wheelandev}
\bibfield{author}{\bibinfo{person}{S Wheelan}.}
  \bibinfo{year}{2005}\natexlab{}.
\newblock \bibinfo{booktitle}{{\em Group processes: {A} developmental
  perspective\/} (\bibinfo{edition}{2} ed.)}.
\newblock \bibinfo{publisher}{Allyn and Bacon}, \bibinfo{address}{Boston}.
\newblock
\showISBNx{0-205-41201-7}


\bibitem[\protect\citeauthoryear{Wheelan, Davidson, and Tilin}{Wheelan
  et~al\mbox{.}}{2003}]%
        {wheelan2003}
\bibfield{author}{\bibinfo{person}{Susan Wheelan}, \bibinfo{person}{Barbara
  Davidson}, {and} \bibinfo{person}{Felice Tilin}.}
  \bibinfo{year}{2003}\natexlab{}.
\newblock \showarticletitle{Group Development Across Time: {R}eality or
  Illusion?}
\newblock \bibinfo{journal}{{\em Small group research\/}} \bibinfo{volume}{34},
  \bibinfo{number}{2} (\bibinfo{year}{2003}), \bibinfo{pages}{223--245}.
\newblock


\bibitem[\protect\citeauthoryear{Yu}{Yu}{2014}]%
        {seovercoming}
\bibfield{author}{\bibinfo{person}{L. Yu}.} \bibinfo{year}{2014}\natexlab{}.
\newblock \bibinfo{booktitle}{{\em Overcoming Challenges in Software
  Engineering Education: Delivering Non-Technical Knowledge and Skills}}.
\newblock \bibinfo{publisher}{IGI Global}, \bibinfo{address}{Hershey,
  Pennsylvania}.
\newblock
\showISBNx{9781466658011}


\end{thebibliography}

\end{document}